\documentstyle[11pt,newpasp,twoside,epsf]{article}
\def\edcomment#1{\iffalse\marginpar{\raggedright\sl#1\/}\else\relax\fi}
\reversemarginpar

\begin{document}
\title{Mid-IR Imaging of AGB Circumstellar Envelopes}
\author{Massimo Marengo, Giovanni G. Fazio, Joseph L. Hora}
\affil{Harvard-Smithsonian CfA, 60 Garden St., Cambridge, MA 02138, USA}
\author{William F. Hoffmann}
\affil{Steward Observatory, University of Arizona, Tucson, AZ 85721, USA}
\author{Aditya Dayal}
\affil{IPAC/Caltech, 770 S. Wilson Ave., Pasadena, CA 91125}
\author{Lynne K. Deutsch}
\affil{Boston University, 725 Commonwealth Ave., Boston, MA 02215, USA}

\begin{abstract}
AGB stars, the precursors of Planetary Nebul\ae, exhibit high rates of 
mass loss and eject material in the form of a slow (10-20 km/s), dusty 
molecular wind. The general belief that the dust component of AGB 
circumstellar envelopes have smooth density profiles and spherical 
symmetry have recently been shaken by new high resolution images,
showing that clumpy and asymmetric structures, analogous to the ones 
observed in Planetary Nebul\ae, can be present even before the end of
the AGB. To illustrate how and when these structures appear, and
possibly to address the question of how they may shape the subsequent
evolution of these systems, we have started a campaign of mid-IR high
spatial resolution imaging of a selected sample of AGB targets. We
want here to illustrate our technique, developed for the mid-IR camera 
MIRAC3, and to show the first results obtained at the NASA Infrared
Telescope Facility.
\end{abstract}

\section{Looking for asymmetrical progenitors of PN}

Intermediate and low mass stars (1--8 M$_\odot$) are characterized, in 
the phase known as Asymptotic Giant Branch (AGB), by the formation of
an optically opacque circumstellar envelope of gas and dust, which
will later evolve into a Planetary Nebula. The detailed physical
processes involved in this phenomena are still uncertain, but there
are growing evidences that they are connected to radial stellar
oscillations and non uniform density distributions (Lebertre \&
Winters 1998; Fleisher et al. 1992).

Recent observations at different wavelengths support the idea that
these inhomogeneities can propagate in the circumstellar envelope,
giving rise to structures with strong deviations from spherical
symmetry. Clumpy structures in the dust forming regions of the
C-rich AGB star IRC $+$10216 were found by 
near-IR masking and speckle interferometry at Keck and SAO
telescopes (Monnier et al. 1997, Weigelt et al. 1998). A sequence of
detached dust shells were also found around this source by deep
optical imaging (Mauron \& Huggins 1999), suggesting a complex mass
loss history similar to the one that characterized the more evolved
post-AGB ``Egg Nebula'' (Sahai et al. 1998), or the O-rich Mira R Hya
(Hashimoto et al. 1998). All these observations
suggest the idea that the asymmetry observed in many PN already 
starts during the AGB, where it shapes the evolution of the circumstellar
envelope towards the Planetary Nebula phase.

Mid-IR is the ideal spectral range to image the spatial distribution
of dust in the circumstellar environment of AGB stars, and provides an 
effective diagnostic tool to derive the physical and chemical
parameters of circumstellar dust (Marengo et al. 1999). The 
availability of a large sample of spatially resolved AGB envelopes in
the mid-IR would be essential to improve our knowledge of the mass
loss processes at the end of the AGB, in search for departures from
spherical symmetry and the ``steady mass loss'' $1/r^2$ radial density 
profile of the stellar outflow.

\section{Modelling the envelope emission}

In most cases, the thermal radiation coming from AGB circumstellar
envelopes is too faint to be detected around the bright AGB star at
the center of the system. Furthermore, only the nearest sources are
extended enough to be spatially resolved with the angular resolution
provided by available IR telescopes.

For these reasons we have compiled our target list fitting all the
available IRAS Low Resolution Spectra (LRS) of AGB sources with the
public domain radiative transfer code DUSTY (Ivezi\'c \& Elitzur
1997). Each computed radial brightness profile was then transformed into
a two dimensional image, convolved with the instrumental PSF, and
resampled into the MIRAC3 final image array. Gaussian
noise was finally added to produce a peak S/N of $\sim$1,000, as
expected for the real observations. Only sources
showing a detectable excess emission above the instrumental Point
Spread Function (PSF), in a minimum area of 6--8 arcsec in diameter,
were selected for observations.

\section{Diffraction limited imaging in the mid-IR}

The circumstellar dust emission predicted by our models is typically
characterized by a compact component, only partially resolved as an
enlarged (in terms of the FWHM) PSF, plus a faint ``halo'' that can be 
separated by the ``wings'' of the PSF only when the S/N is of the 
order of $\sim$1,000, or larger. For a positive detection of these two 
components is necessary to maximize the achievable angular resolution
and sensitivity. To meet these requirements,
we are using a technique based on the ``fast readout'' mode of the
camera MIRAC3, that allows the acquisition of frames with very short
integration times (0.1-0.2 sec), capable of ``freezing'' the
atmospheric seeing.

The source is imaged with a standard nodding and chopping technique,
in order to remove the background signal, and dithered on the array to 
obtain for each beam a number of images that can be of the order of
$\sim$500. Each image is then rebinned on a sub-pixel grid to
increase the PSF sampling, shifted and coadded. 

To maximize the sampling of the dithered images without degrading the
angular resolution, we have adopted the ``drizzling'' method developed by
Fruchter \& Hook (1997), whenever the number of single frames
available for the coadding is sufficient to provide an uniform
distribution of the dithered ``drops'' on the final image (tipically
200-250 frames for each beam). As shown in Fig.~1, this
technique allow to increase the S/N reducing the PSF FWHM,
producing images that are virtually diffraction limited,
and largely independent from the atmospheric seeing.

\begin{figure}
\plotone{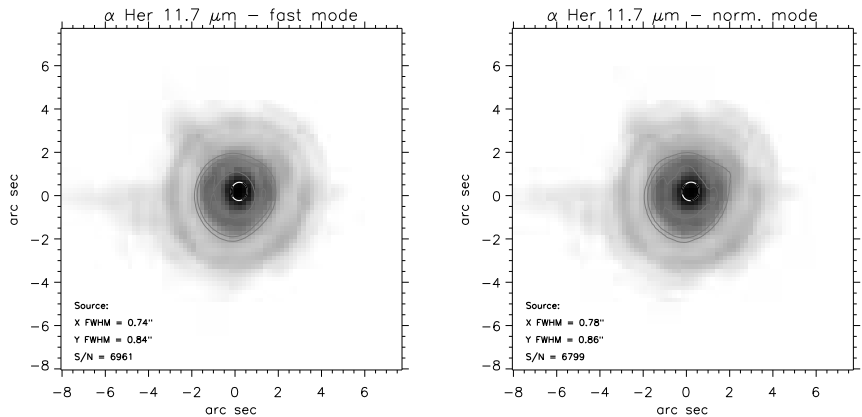}
\caption{The reference star $\alpha$ Her at 11.7 $\mu$m. The image
on the left was obtained using MIRAC3 fast mode, coadding about 1,600
single exposures of 0.17 sec (272 sec total); on the right is the
coadding (without drizzling) of 80 exposures of 10 sec each (800 
sec total). Note the smaller FWHM of
the fast mode image, and its higher S/N (despite the shorter total
integration time).}
\end{figure}

\begin{figure}[ht]
\plotone{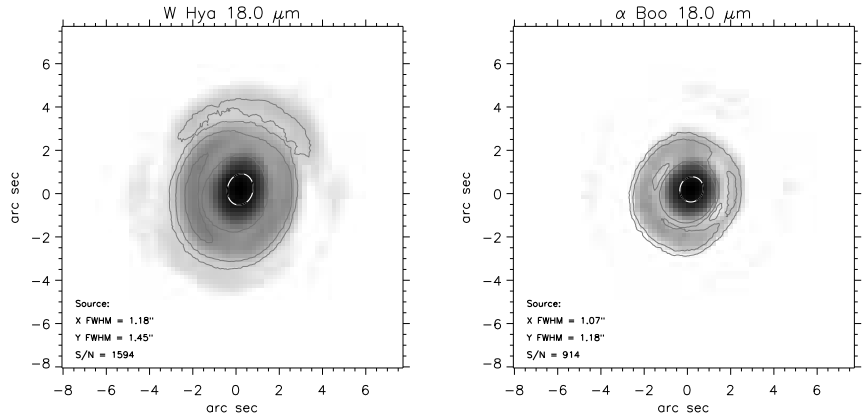}
\caption{18 $\mu$m images of the O-rich semiregular variable W Hya,
and the bright IR standard star $\alpha$ Boo as PSF
reference. Contours levels indicate 0.01, 0.02, 0.05 and 0.5
of the source and reference maximum.}
\end{figure}

One of the most extended sources observed in June 1999 run with MIRAC3 
at IRTF is the O-rich semiregular variable W Hya. We present here its
18 $\mu$m image, compared with the reference star $\alpha$ Boo
(Fig.~2). Note the much larger FWHM of the AGB source, and its oval
shape elongated in the N-S direction, compared to the more compact
and symmetric image of the reference star. In June 1999 run, a total 
of 10 AGB sources where observed at 8.8, 9.8, 11.7, 12.5 and 18 $\mu$m; 
the analysis of the collected data is currently in progress, and further 
observing runs are scheduled for IRTF and MMT.

\end{document}